\documentclass[12pt]{article}
\usepackage{color}
\usepackage{graphicx}

\definecolor{violet}{rgb}{0.4,0,0.6}
\definecolor{vert}{rgb}{0,0.6,0.2}
\definecolor{navy}{rgb}{0.0,0.0,0.4}

\def\colbrun#1{\textcolor[named]{Brown}{#1}}

\def\spose#1{\hbox to 0pt{#1\hss}}\def\lta{\mathrel{\spose{\lower 3pt\hbox
{$\mathchar"218$}}\raise 2.0pt\hbox{$\mathchar"13C$}}}  \def\gta{\mathrel
{\spose{\lower 3pt\hbox{$\mathchar"218$}}\raise 2.0pt\hbox{$\mathchar"13E$}}}
\def\Libra{\spose {--} {\cal L}}

\font\sixrm=cmr6

\def\sigme{{\color{vert}\sigma}} \def\xe{{\color{vert}\xi}}
\def\ete{{\color{vert}\eta}}\def\Ke{{\color{vert}K}}
\def\delte{{\delta}}

\def\dP{\spose {\lower 2.0pt\hbox{$_{_\Gamma}$} } {\delte}}
\def\dL{\spose {\lower 5.0pt\hbox{\sixrm \color{vert}L} } {\delte}}
\def\dE{\spose {\lower 5.0pt\hbox{\sixrm E} } {\delta}}

\def\xiI{\spose {\raise 3.0pt\hbox{$\, \acute{\ }$}} {\xe}}
\def\xiJ{\spose {\raise 3.0pt\hbox{$\, \grave{\ }$}} {\xe}}
\def\deI{\spose {\raise 3.0pt\hbox{$\, \acute{\ }$}} {\delte}}
\def\deJ{\spose {\raise 3.0pt\hbox{$\, \grave{\ }$}} {\delte}}

\def\rmA{ {_{\rm \color{blue}A}}}  
\def\rmB{ {_{\rm \color{blue}B}}}

\def\thetar{ {\color{red}{\Theta} }}
\def\varthetar{ {\color{red}{\vartheta} }}
\def\omegar{ {\color{red}{{\Omega}} } }
\def\varpir{ {\color{red}{\varpi} } }
\def\varpirIJ{\spose {\raise 3.0pt\hbox{$\, \acute{\ }\grave{\ }$}}{\varpir}}

\def\cur{ {\color{red}{c}} }

\def\calI{ {\color{red}{\cal I}} }
\def\calH{ {\color{red}{\cal H}} }
\def\calL{ {\color{red}{\cal L}} }    \def\Lred{ {\color{red} L} }
\def\calK{ {\colbrun{\cal K}} }

\def\Pred{ {\colbrun{p}} }
\def\Pired{ {\colbrun{\pi}} }
\def\pred{ {\colbrun{p}} }

\def\Cred{ {\color{red}C} }           \def\mred{ {\color{red}m} }
\def\Tred{ {\color{red}T} }
\def\Ured{ {\color{red}{U}} }
\def\Tred{ {\color{red}{T}} }
\def\vlred{ {\color{red}{c_L}} }
\def\vered{ {\color{red}{c_E}} }

\def\glambda{ {\color{violet}\lambda} }
\def\gmu{ {\color{violet}\mu} }

\def\ii{ {\color{vert}i} }            \def\ji{ {\color{vert}j} }
\def\sigme{ {\color{vert}\sigma} }    \def\nable{ {\color{vert}\nabla} }
\def\ete{ {\color{vert}\eta} } \def\gamme{ {\color{vert}\gamma} }
\def\perpe{ {\color{vert}\perp} }
\def\Gamme{ {\color{vert}\Gamma} }
\def\he{ {\color{violet}h} }

\def\vq{ {\color{violet} q} } \def\vw{ {\color{violet} w} }
\def\vphi{ {\color{violet} \varphi} }\def\vpsi{ {\color{violet} \psi} }
\def\kap{ {\color{violet} \kappa_{_0}} }

\def\bb{$}  \def\fb{$ }
\def\be{\begin{equation} }
\def\fe{\end{equation}}
\def\ee{\end{equation}}
\newcommand{\ba}{\begin{eqnarray}}
\newcommand{\ea}{\end{eqnarray}}

\begin{document}
\bigskip
\hskip 4.8in\vbox{\baselineskip12pt \hbox{LPT-ORSAY 03-52} }

\bigskip
\bigskip
\bigskip

\begin{center}
{\Large \textcolor{red}{\Large Symplectic structure  for elastic
and chiral conducting cosmic \\[0.4cm]
string models}}
\end{center}
\bigskip
\bigskip
\bigskip
\centerline{\large
\textcolor[named]{ForestGreen}{B.~Carter}$^\sharp$ and
\textcolor[named]{ForestGreen}{D.~A.~Steer}$^\flat$}
\bigskip
\bigskip
\centerline{ {$^\sharp$}LuTh, Observatoire de Paris-Meudon, 92195
Meudon, France.} \bigskip\centerline{{$^\flat$}Laboratoire de
Physique Th\'eorique\footnote{Unit\'e Mixte de Recherche du CNRS
(UMR 8627).} } \centerline{B\^at.\ 210, Universit\'e Paris XI,
91405 Orsay Cedex, France} \centerline{and} \centerline{
F\'ed\'eration de recherche APC, Universit\'e Paris VII,}
\centerline{2 place Jussieu, 75251 Paris Cedex 05, France.}
\bigskip
\bigskip
\centerline{July 2003}

\vskip 3 cm

\begin{abstract}
This article is based on the covariant canonical formalism and
corresponding symplectic structure on phase space developed by
Witten, Zuckerman and others in the context of field theory. After
recalling the basic principles of this procedure, we construct the
conserved bilinear symplectic current for generic elastic string
models.  These models describe current carrying cosmic strings
evolving in an arbitrary curved background spacetime.  Particular
attention is paid to the special case of the chiral string for
which the worldsheet current is null.  Different formulations of
the chiral string action are discussed in detail, and as a result
the integrability property of the chiral string is clarified.
\end{abstract}

\vfill\eject

\noindent
{\bf 1. Introduction}
\medskip

Following a preliminary study~\cite{Ca03} whose application was
restricted to the non-conducting Nambu Goto (NG) strings, in this
article we focus on conducting cosmic string models and apply the
general principles of covariant canonical variational analysis.
As emphasised by Witten, Zuckerman, and
others~\cite{Wi86,CrWi87,Zu87,So94,CF98,Nu00,Ro02} in the context
of relativistic field theories, the potential utility of this
covariant analysis is as a starting point for covariant
quantization.
Despite that, our aim is not to quantize.  Rather, the
culmination of this analysis is the construction of a
corresponding symplectic structure which is locally representable
as a current.  This symplectic current --- which should not be
confused with the world-sheet current of the conducting string ---
is defined as an antisymmetric bilinear functional of a pair of
independent perturbations, and it is conserved whenever both
perturbations are on shell (in the sense of satisfying the
relevant dynamical field equations).

The task of extending such analysis from ordinary fields to branes
(meaning systems with support confined to a lower dimensional
worldsheet) was recently taken up by
Cartas-Fuentevilla~\cite{CF02,CF02b}. The necessary analysis is
facilitated by the relatively new
development~\cite{Ca93,BaCa95,BaCa00} of suitably covariant
methods of geometrical analysis, which have already been shown to
be far more efficient than the more cumbersome (and error prone)
frame dependent methods used in earlier work for treating other
problems, such as the divergences arising from self
interaction~\cite{Ca97,CaBa98,CaBaUz02}.

The preceding article~\cite{Ca03} demonstrates the agreement of
the canonical approach, as developed by
Cartas-Fuentevilla~\cite{CF02,CF02b}, with the result of an
earlier and rather different approach~\cite{Ca93} to the
construction of the symplectic surface current.  Both these
analyses focused on strings with no internal physical structure on
the worldsheet, that is on actions of the simple Dirac-Nambu-Goto
type which are proportional to the worldsheet surface measure.

The present work deals with the more general case of elastic string
models of the kind~\cite{Ca89} appropriate for macroscopic applications,
such as the strings used for musical instruments since the time
of Pythagorus, and in particular for the macroscopic description
of the effect in cosmic strings of mechanisms of the various (fermionic and
bosonic superconducting) kinds originally proposed by Witten~\cite{Wi85}.
The earliest work on the consequences of the Witten mechanism emphasised
effects due to electromagnetic coupling, but it was recognised by
Davis and Shellard~\cite{DaSh88} that the most important consequence would
be the
formation of vortons, meaning centrifugally supported
equililibrium configurations of loops in which effects of electromagnetic
coupling (if present at all) are relatively unimportant.  While it is
also possible to have non-circular vorton configurations~\cite{CaMa93},
the simplest possibility is that of circular configurations.  In this case
(provided electromagnetic effects are absent or negligible) both
the static equilibrium and full
dynamical evolution is particularly amenable to an exact mathematical
treatment: this was provided by a recent study~\cite{CaPeGa97}
whose notation scheme will be followed here.

This paper is set up in the following way.  In section 2 we review
the generic elastic string model and the limit case of the
chiral model, and derive the
relevant equations of motion.  Different formulations of the
chiral string action are discussed in detail in section 2a, and as
a result we are able to reinterpret more simply the integrability property of
the chiral string.  The purpose of section 3 is to summarize the
basic steps of the canonical analysis of Witten and others, and
this is done in the context of a general worldsheet action
density.  In section 4 we discuss in which reference system the
symplectic current will be evaluated; the result for the general elastic
string is given in section 5, and for the
chiral string in section
6.  Finally this paper contains an appendix in which we use
Dirac's Hamiltonian method for constrained systems in order to
write the chiral string action in a linearised Polyakov-like form.

Since there are many different indices to keep under control in
this analysis, we have tried to clarify the presentation by using
a colour scheme.  Black indices refer to spacetime quantites;
green quantities are geometrical, describing the embedding of the
brane (in this case a string) in the background spacetime; and
blue indices on a vector run over both spacetime and internal
indices. We have decided to write all physical quantities
(invariant under gauge transformations and rescalings) in red;
generalised momenta are written in brown; and quantities in purple
are dynamical (generally gauge or normalisation dependent)
variables.

\bigskip\noindent
 {\bf 2. Elastic and chiral string models}
\medskip

The string models we consider are governed by an action integral
of the form
\be
{\calI}=\int \Lred \Vert\gamme\Vert^{1/2} \, {\rm d}^2\sigme\,
,\label{1-}
\ee
over a supporting worldsheet with internal
coordinates  $\sigme^\ii$ $(\ii=0, 1)$ and induced metric
\bb
{\gamme}_{\ii\ji}=g_{\mu\nu}x^\mu_{,\ii}x^\nu_{,\ji},
\fb
in a
background with coordinates \bb x^\mu\, ,\fb \bb(\mu=0, 1, ...\,
d)\, ,\fb \bb (d\geq 2)\ \fb and (flat or curved) space-time
metric \bb g_{\mu\nu}.\fb  In the NG case, the scalar Lagrangian
action density \bb\Lred \fb is just a constant. For the more
general category~\cite{CaPeGa97} of elastic string models
considered here, \bb\Lred \fb depends on the magnitude of the
gradient of a freely variable phase field \bb\vphi.\fb  More
explicitly the equation of state is $\Lred\{\vw\}$ where
\be
\vw=\vpsi^2\gamme^{\ii\ji}\vphi_{,\ii}\vphi_{,\ji}\, \label{3-}
\ee
with \bb\vpsi\fb a normalisation parameter with fixed value given
\be \vpsi^2=\kap\, .\label{4-}
\ee
The constant \bb\kap\fb may be fixed according to convenience
without loss of generality of the model: the standard
convention~\cite{CaPeGa97}, however, is to choose \bb\kap\fb such
that the derived quantity \bb\calK\fb defined by
\be \calK^{-1}=-2{{\rm d}\Lred\over {\rm d}\vw} \,\, ,\label{5-}\fe
tends to unity in the null limit $\vw \rightarrow 0$ (see
equation \ref{15-} and section 2a).

For any such elastic string action there is a corresponding {\it
chiral} string action which is obtained by relaxing the condition
that \bb\vpsi\fb is fixed, and letting it instead be a freely
varying auxiliary field.  On application of the variation
principle, it follows that the on shell configurations satisfy the
same dynamical equations as in the corresponding elastic model,
though there is now the further constraint that $\vw=0$. It can be
verified that this null constraint is consistent with the
dynamical equations:
if $\vw=0$ at an initial time, it will automatically be preserved
by the evolution of the system.

This chiral string is of interest~\cite{MaSh98} since in a number
of different and cosmologically relevant
cases~\cite{DaDaTr97,CaDa00}, Witten's fermionic zero mode
mechanism indeed gives rise to purely left (or purely right)
moving modes and hence to a null current.  Furthermore it was
recently recognized~\cite{CaPe99,DaKiPiSt00,BlOlVi01} that the
chiral string equations of motion are exact integrable in a flat
background (see also further comments on this in section 2).  This
has led to a number of analyses of the cosmological consequences
of chiral cosmic strings~\cite{DaKiPiSt00,St01}. The chiral string can also be
thought of as a useful approximation for the treatment of nearby
solutions of the less tractable generic elastic class.

As in the simple NG case ($\vphi =0$), the full set of dynamical
equations for the generic elastic case is given by the local
surface energy momentum conservation equation, which takes the
standard form
\be \overline\nable_{\!\mu}\Tred^{\mu\nu}=0\, .\label{7-}\fe
Here the surface covariant differentiation operator \bb
\overline\nable\fb is defined in terms of the fundamental tensor
\bb \ete^{\mu \nu}\fb
by
\be \overline\nable_{\!\mu}=\ete_\mu^{\,\nu}\nabla_\nu\, ,\hskip 1 cm
\ete^{\mu\nu}=\gamma^{\ii\ji}x^\mu_{,\ii}x^\nu_{,\ji}
\, ,\label{8-}\fe
and the surface stress momentum energy density  is defined by
\be
\Tred^{\mu\nu}=2\Vert\gamme\Vert^{-1/2}{\partial\big(\Lred
\Vert\gamme\Vert^{1/2} \big)\over \partial g_{\mu\nu}}\,
.\label{9-}
\ee
The motion of the worldsheet is governed by the
orthogonally projected part of (\ref{7-}) which takes the form
\be
\Tred^{\mu\nu}\Ke_{\mu\nu}{^\rho}=0 \, ,\label{10-}
\ee
where $\Ke_{\mu\nu}{^\rho}$ is the second fundamental tensor as
defined~\cite{Ca89} by
\be
\Ke_{\mu\nu}{^\rho}= \ete^\sigma_{\,\nu}\overline
\nable_{\!\mu}\ete^\rho_{\,\sigma}\, .\label{11-}
\ee
For the NG string \bb \Tred^{\mu\nu}\fb is simply proportional to
the first fundamental tensor $\ete^{\mu\nu}$, and the extrinsic
evolution equation (\ref{11-}) by itself constitutes the complete
set of dynamical equations. However, in the elastic case, the
surface stress energy tensor is slightly more complicated;
\be
\Tred^{\mu\nu}=\Lred\ete^{\mu\nu}+\calK\,\cur^\mu \cur^\nu \,
,\label{12-}
\ee
involving a surface current \bb\cur^\mu \fb given by
\be
\cur^\mu={\vpsi\over\calK}\overline\nable^\mu\vphi=
{\vpsi\over\calK}x^\mu_{,\ii}\gamma^{\ii\ji}\vphi_{, \ji} \,
.\label{13-}
\ee
Now the full set of dynamical equations is given by the extrinsic
evolution equation (\ref{10-}), together with the surface current
conservation law
\be \overline \nable_{\!\mu}\cur^\mu=0 \, .\label{14-}\fe

In the chiral case these equations also apply, though they must be
supplemented by the constraint $\vw=0$, which means that the
current \bb\cur^\mu \fb is null. In this null current limit, the
Lagrangian \bb\Lred\fb tends to a fixed value specified by the
relevant Kibble mass scale $\mred$ say, and for the standard
choice of \bb\kap\fb the quantity $\calK$ tends to unity according
to the specifications
\be
\Lred\{0\}=-\mred^2\, ,\hskip 1 cm \calK\{0\}=1\,  \label{15-}
\ee
(see section 2a). Thus in the chiral case the stress energy tensor
will take the form
\be  \Tred^{\mu\nu}=-\mred^2\ete^{\mu\nu}+\cur^\mu \cur^\nu
\, ,\label{16-}
\ee
subject to the nullity condition
\be
\cur^\mu \cur_\mu=0\, ,\label{17-}
\ee
where the current is given simply by
\be
\cur^\mu=\vpsi\,\overline\nable^\mu\vphi= \vpsi\,
x^\mu_{,\ii}\gamma^{\ii\ji}\vphi_{, \ji} \, .
\label{18-}
\ee
Recall that whereas \bb\vpsi\fb is held constant in the analogous
formula (\ref{13-}) for the elastic case, it may vary in
(\ref{18-}) for the chiral case.  However, the on shell
variations of \bb\vpsi\fb are severely restricted since on a two
dimensional world sheet, the nullity condition (\ref{17-}) (i.e.\
$\vw=0$), automatically imposes that the phase field must satisfy
the harmonicity condition
\be \big(\Vert\gamme\Vert^{1/2} \gamme^{\ii\ji}
\vphi_{\, ,\ii}\big){_{, \ji}}=0 \,  .\label{19-}
\ee
It can thus
be seen from the internal coordinate version of the current
conservation law (\ref{14-}), namely
\be
\big(\Vert\gamme\Vert^{1/2} \gamme^{\ii\ji}\vpsi \vphi_{\,
,\ii}\big){_{, \ji}}=0 \, ,\label{20-}
\ee
that the auxiliary field
$\vpsi$ must satisfy the dynamical evolution condition
\be \gamme^{\ii\ji}\vphi_{,\ii}\vpsi_{\, ,\ji}=0 \, ,\label{21-}\fe
which implies that \bb\vpsi\fb will be restricted on shell to be a
function just of \bb\vphi\fb.

\bigskip\noindent
 {\bf 2a. Further properties of chiral strings and different actions}
\medskip

Whereas different equations of state for \bb\Lred\{\vw\}\fb give
qualitatively different elastic string models, the chiral model on
the other hand is {\it unique} (just as in the NG case), modulo
the choice of the fixed mass scale $\mred$.  Thus there is no loss
of generality in taking the chiral string equation of state to
have the simple linear form originally proposed by
Witten~\cite{Wi85}, namely
\be
\Lred=-\mred^2-{_1\over^2}\vw \ \ \Rightarrow \ \ \calK=1 \,
.\label{22-}
\ee
However, it is instructive to consider the many different allowed
alternatives which may, for instance, make the integrability
property of chiral string mentioned above more transparent.
Indeed one can see that the equations of motion (\ref{17-}) and
(\ref{20-}) as well as the stress energy tensor (\ref{16-}) for
the chiral string can be obtained from any action of the form
\be
\Lred=-\mred^2-{_1\over^2}\vw + a_2 \vw^2 + a_3 \vw^3 + \ldots \,
,\label{22a-}
\ee
where the $a_n$ ($n=2,3,\ldots$) are arbitrary numerical
constants. These additional terms do not contribute on shell since
in the equations of motion they lead to terms proportional to
$\vw^{n-1}$ which vanish in the chiral limit $\vw=0$. Off-shell,
however, they are important. As we now explain, a particularly
useful version of this action for many reasons is the non-linear
form
\be
\Lred=-\mred^2\big(1+\vw/\mred^2\big)^{1/2} \ \ \Rightarrow \ \
\calK=\big(1+\vw/\mred^2\big)^{1/2}\, , \label{23-}
\fe
which we shall often call the `square root' action.

First consider action~(\ref{23-}) for arbitrary (generally
non-zero) $\vw$. From the corresponding stress-energy tensor
it follows that the energy density $\Ured$ and
tension $\Tred$ satisfy
\be
\Ured \Tred ={\rm const}.
\ee
Hence the two sound speeds $\vered$ and $\vlred$ (respectively
the transverse ``wiggle'' and longitudinal ``woggle'' sound speeds
\cite{Bwiggle}) defined by
\be
\vlred^2 = \frac{\Tred}{\Ured} \qquad , \qquad
\vered^2 = -\frac{d\Tred }{d\Ured }
\ee
are equal for action (\ref{23-}).  (For a NG string, $\vered=1$
whilst $\vlred$ has no analogue.) This special case $\vlred = \vered$
characterises the transonic elastic string model that was already
known to be integrable for any $\vw$ \cite{Ca90}. Thus, turning
now to the chiral case, once one has realised the equivalence of
actions (\ref{22-}) and (\ref{23-}) it comes as no surprise that
the chiral action is also integrable (as was first proved in
reference \cite{CaPe99} though on the basis of the action in the
form (\ref{22-})).

A second feature of the action corresponding to the Lagrangian given in
(\ref{23-}) is that it can viewed
as a Kaluza-Klein projection of a NG action in a space-time of
one higher dimension \cite{Bbluebook,Cordero}.  Indeed
\be
{\calI}=-\mred^2 \int {\rm d}^2\sigme  \Vert\gamme\Vert^{1/2}
\big(1+\vw/\mred^2\big)^{1/2}
=-\mred^2 \int {\rm d}^2\sigme  \Vert \Gamme\Vert^{1/2}
\, ,\label{KK-}
\ee
where ${\Gamme}_{\ii \ji}$ is now the projection of the $(d+2)$-dimensional
metric
\be
{G}_{\rmA \rmB} = \left(
\begin{array}{cc}
g_{\mu \nu} & 0 \\
0 & \vpsi^2/\mred^2 \\
\end{array}
\right)
\ee
with the brane embedding is given by
\be
\vq^{\rmA} = \left(
\begin{array}{c}
x^{\mu}(\sigme^\ii) \\
\vphi(\sigme^\ii) \\
\end{array} \right),
\label{qadef}
\ee
so that $\Gamme_{\ii \ji} = \partial_{\ii} \vq^{\rmA} \partial_{\ji} \vq^{\rmB}  {G}_{\rmA \rmB}$.

Finally it is interesting to observe that the action coming from
(\ref{23-}) can be put into a Polyakov-like form --- that is, a
form which depends linearly on $\gamme_{\ii \ji}$ --- but this is
not possible for the action coming from (\ref{22-}): in the
appendix we demonstrate this using Dirac's Hamiltonian formalism
for contrained systems \cite{Dirac,Hansen} and also discuss the
algebra of constraints. As expected, the result for the Polyakov
action is
\ba
{\calI}
& = & -\mred^2 \int {\rm
d}^2\sigme  \Vert\gamme\Vert^{1/2} \big(1+\vw/\mred^2\big)^{1/2}
\label{28a-}
\\
&=&-\frac{\mred^2}{2} \int {\rm d}^2\sigme
\Vert\he\Vert^{1/2} \he^{\ii \ji} \left(g_{\mu \nu} \partial_{\ii}
x^{\mu} \partial_{\ji}x^{\nu} + \frac{\vpsi^2}{\mred^2}
\partial_{\ii}\vphi \partial_{\ji}\vphi \right)
\label{Lpolyfinalt}
\ea
where $\he^{\ii \ji}$ is now an auxilliary metric field (again
see the appendix). Note that this Polyakov action is
reparametrisation and scale invariant so that coordinates can be
chosen such that $\he_{\ii \ji} = \eta_{\ii \ji}$ where $\eta_{\ii \ji}$
is the Minkowski metric. As for the Nambu string \cite{Green}, this could
be a starting point to discuss a possible quantization of the
chiral string~\cite{Marti}.

\bigskip
\noindent
{\bf {3. General canonical symplectic structure}}
\medskip

Our aim is to subject the elastic and chiral string models to a
canonical analysis of the kind described in the preceding
article~\cite{Ca03}.  However, before doing so, we summarise in
this section the main features of the canonical analysis.  The
following discussion is general and is not restricted to a
particular string model.

Consider the general case of a worldsheet action density
\be
{\calL}=\Vert{\ete}\Vert^{1/2}{\Lred}\, ,\label{28-}
\ee
in which the Lagrangian density $\Lred$ depends on a set of field
components $\vq^\rmA$ and of their surface deriatives
$\vq^\rmA_{\, ,\ii}
=\partial_\ii\vq^\rmA=\partial\vq^\rmA/\partial\sigme^\ii$ where
the $\sigme^{\ii}$ are the worldsheet coordinates.  In this
discussion we do not restrict the worldsheet dimension to 2, so
that the formalism presented here will be applicable not just to
strings but also to higher dimensional branes. The field variables
$\vq^\rmA$ can be of internal or external kind, the most obvious
example of the latter kind being the background coordinates \bb
x^\mu\fb themselves.  (In the case of the elastic string $\vq^\rmA
=(x^{\mu},\vphi)$.)

Subject to the understanding that the internal coordinates are held
fixed,
\be \delte\sigme^\ii=0\, ,\label{29-}\fe
the generic action variation,
 \be \delte{\calL}= {\calL}_{\!\rmA}\delte\vq^\rmA +
\pred_{\!\rmA}^{\, \ii}\delte\vq^\rmA_{\, ,\ii}\, ,\label{30-}\fe
specifies a set of partial derivative components
\bb{\calL}_{\!\rmA} \fb and an  associated set of generalised
momentum components ${\pred}_{\!\rmA}^{\, \ii}$, which can be used
to construct a corresponding pseudo-Hamiltonian scalar density
\be
\tilde{\calH}= {\pred}_{\!\rmA}^{\, \ii}\vq^\rmA_{\, ,\ii}
-\calL\, .\label{31-}
\ee
(The covariance of such a pseudo-Hamiltonian distinguishes it from
the ordinary kind of Hamiltonian, which depends on the
introduction of some preferred time foliation as for instance the
appendix of this article, equation (\ref{canonH}).)

According to the variational principle, the dynamically admissible
``on shell'' configurations are those characterised by the
vanishing of the Eulerian derivative given by
\be
\frac{\delte \calL}{\delte\vq^\rmA}=
\calL_{\!\rmA}-{\pred}_{\!\rmA\, ,\ii}^{\, \ii}\, .\label{32-}
\ee
For an on-shell configuration, i.e. when the dynamical equations
\be
\frac{\delte \calL}{\delte\vq^\rmA}=0\, ,\label{33-}
\ee
are
satisfied, the pseudo-Hamiltonian variation will take the form
\be
\delte \tilde{\calH}=\vq^\rmA _{\, ,\ii}\delte{\pred}_{\!\rmA}^{\,
\ii} -\pred_{\!\rmA\, ,\ii}^{\, \ii}\delte\vq^\rmA      \,
\label{34-}.
\ee
Thus the Lagrangian variation can be written as a pure surface
divergence
\be \delte\calL=\varthetar^\ii_{,\ii}\,
\label{35-}
\ee
where $\varthetar^\ii$ is the generalised Liouville 1-form (on the
configuration space cotangent bundle) defined by
\be
\varthetar^\ii=\pred_\rmA^{\, \ii}\delte \vq^\rmA\,
.\label{36-}
\ee
Equation (\ref{35-}) shows that the Liouville 1-form is
interpretable as a surface current that will be conserved (in the
sense of having vanishing surface divergence) {\it provided} it is
constructed from a perturbation that generates a local symmetry of
the Lagrangian density, i.e. such that \bb \delte\calL=0\fb.  In
the general case it is not conserved.

We can go on to construct a surface current that will always be
conserved when the relevant dynamical equations are satisfied.
This is done by taking the exterior differential of the Liouville
form, i.e.\ by evaluating the commutator of a pair of successive
independent variations, in the manner described in detail in the
preceding article~\cite{Ca03}. This exterior variation procedure
provides us with the closed (since manifestly exact) symplectic
2-form expressible as
\be \varpir^\ii=\delte\wedge\varthetar^\ii=\delte
\pred_\rmA^{\, \ii}\wedge\delte \vq^\rmA\, \label{37-}
\ee
where we have used the wedge symbol $\wedge$ to indicate
antisymmetrisation with respect to the two independent variations
involved. (Many authors prefer to use an extreme kind of
abbreviation scheme in which the wedge symbol is omitted,  but ---
as discussed in~\cite{Ca03} --- the use of such ultra-concise
notation can lead to confusion in cases involving symmetrisation
as well as antisymmetrisation). It is easy to verify that whenever
both perturbations satisfy the relevant perturbed field equations
\be
\delte\left( \frac{\delte \calL}{\delte \vq^\rmA}\right)=0\, ,
\label{38-}
\ee
the symplectic 2-form will be interpretable as a conserved
worldsheet current in the sense that it will satisfy
\be
\varpir^\ii_{\, ,\ii}=0\, .
\label{39-}
\ee

We now follow the strategy of~\cite{Ca93} and (as far as possible)
work with quantities that are purely tensorial with respect to the
background space.  Hence we translate the surface current
densities, whose components $\varthetar^\ii$ and $\varpir^\ii$
depend on the choice of the internal coordinates $\sigme^\ii$,
into corresponding quantities which have stricly vectorial
background coordinate components.  These are given by
\be
\thetar^\nu=\Vert\gamme\Vert^{-1/2} x^\nu_{\, ,\ii}\varthetar^\ii
\, \qquad \omegar^\nu=\Vert\gamme\Vert^{-1/2} x^\nu_{\,
,\ii}
\varpir^\ii \, .\label{41-}
\ee
The divergence law (\ref{35-}) is now rewritten in terms of the
vectorial version of the Liouville form as
\be
\overline\nable_{\!\nu}\thetar^\nu=\Vert\ete\Vert^{-1/2}\delte
\big(\Lred\,\Vert\gamme\Vert^{1/2}\big)\, . \label{42-}
\ee
Similarly the conservation law (\ref{38-}) simply becomes
\be
\overline\nable_{\!\nu}\omegar^\nu=0\, .\label{43-}
\ee

\bigskip\noindent
{\bf 4. Comoving reference system for the simply elastic case.}
\medskip

We now apply the previous formalism to the elastic string for
which $\vq^{\rmA}= (x^{\mu},\vphi)$, and $\pred^{\rmA} =
(\Pred^{\mu},\Pired)$ where $\Pred_{\mu} = \delta {\calL} /\delta
\dot{x}^{\mu}$, $\Pired = \delta {\calL}  /\delta \dot{\vphi}$.

The meaning of the convention (\ref{29-}) that the local variation
\bb \delta \fb should be evaluated  at a fixed value of the
internal coordinates \bb \sigme^\ii\fb depends on how these
coordinates are chosen. For explicit solutions of the field
equations it may be most convenient to choose cordinates that are
constant along characteristics --- as discussed in section 2a,
this is particularly particularly true in the transonic case for
which the extrinsic and intrinsic characteristics coincide.
However, in the generic elastic case, the most convenient option
is to take the internal coordinates to be comoving with respect to
the intrinsic material structure. For a generic $p$-brane this is
specified in terms of a set of $p$ independent scalar fields; for
an ordinary perfect fluid or more general elastic solid in
4-dimensional spacetime one has $p=3$; whilst in the elastic
string case we are concerned with here one simply has $p=1$ with
the field in question being the phase scalar $\vphi$. Assuming
the absence of singularities where the phase gradient is not just
null but actually vanishes, there will be no loss of generality in
postulating that the internal coordinates are comoving with
respect to $\vphi$.  In other words, we postulate that there is a
vanishing variation, \bb \delte\vphi=0\fb, and hence also
vanishing gradient variation, \bb \delte\vphi_{,\ii}=0\fb with
respect to these coordinates.

Subject to the choice of such a comoving internal reference
system, the only remaining independent field variables in the
elastic string model are the background coordinates $x^\mu$, so
the generic variation (\ref{30-}) will take the specific form
\be
\delte\calL=\calL_\mu\xe^\mu +\pred_\mu^{\ \ii}\,\xe^\mu_{\, ,\ii}
\,
,\label{45-}
\ee
using the notation
\be
\xe^\mu=\delte x^\mu\, \label{46-}
\ee
for the relevant displacement vector. The corresponding expression
for the Liouville form (\ref{36-}) is given by
\be
\varthetar^\ii=\pred_\mu^{\ \ii}\,\xe^\mu \, ,\label{47-}
\ee
while for the symplectic 2-form (\ref{37-}) is given by the
expression
\be
\varpi^\ii=\dP \pred_\mu^{\ \ii}\wedge\xe^\mu\, .\label{48-}
\ee
Here the parallel variation \bb \dP\pred_\mu^{\ \ii}\fb is given
in terms of the simple momentum variation $\delte\pred_\mu^{\
\ii}$ by
\be
\dP\pred_\mu^{\ \ii}=\delte\pred_\mu^{\ \ii} - \Gamma_{\!\mu\
\rho}^{\ \nu}\pred_\nu^{\ \ii}\xe^\rho\, ,
\label{49-}
\ee
where $\Gamma_{\!\mu\ \rho}^{\ \nu}$ are the Riemannian connection
components. In view of the symmetry of the latter it actually
makes no difference whether  \bb\dP\pred_\mu^{\ \ii}\fb or
\bb\delte\pred_\mu^{\ \ii}\fb is used in (\ref{49-}), but use of
the parallel variation is more convenient for our next step, which
is the evaluation of the corresponding background tensorial
formulae.

Since the background coordinate displacement will effect the
Lagrangian only via the change of the induced metric, the resulting
variation will be given simply by
\be
\delte\calL=\frac{1}{2}\Vert\ete\Vert^{1/2}\, \Tred^{\mu\nu}\,\dL
g_{\mu\nu}\, ,
\label{50-}
\ee
where $\Tred^{\mu\nu}$ is the surface stress energy tensor
defined in (\ref{9-}), and $\dL g_{\mu\nu}$ is the Lagrangian
variation of the metric, meaning the change with respect to a
coordinate system that is comoving with respect to the
displacement.  In the absence of any Eulerian variation (meaning
that the spacetime background is held fixed) the Lagrangian
variation is just given by the corresponding Lie derivative:
\be
\dL g_{\mu\nu}=\vec\xe\Libra g_{\mu\nu}
=2\nabla_{\!(\mu}\xe_{\nu)}\, .\label{51-}
\ee
Comparing
(\ref{50-}) with the canonical variation formula (\ref{45-}) the
partial derivatives involved can be read out as
\be
\calL_\mu=\Vert\gamme\Vert^{1/2}\,\Gamma_{\!\mu\ \rho}^{\ \nu}
\Tred_{\!\nu}{^\rho}\, ,\label{52-}
\ee
and
\be
\pred_\mu^{\ \ii}=\Vert\gamme\Vert^{1/2}\,\Tred_{\mu\nu}
\ete^{\ii\ji} x^\nu_{\, ,\ji}\, .
\label{53-}
\ee
It is thus
immediately apparent that the pseudo-Hamiltonian density
(\ref{31-}) will be given by
\be
\Vert\gamme\Vert^{-1/2}\tilde{\calH}=\Tred_{\!\nu}{^\nu}-\Lred \,
,\label{54-}
\ee
and that the vectorial version (\ref{41-}) of the
Liouville current will be given simply by
\be
\thetar^\nu={\Tred}_\mu{^{\nu}}\xe^\mu\, .\label{55-}
\ee

\bigskip\noindent
{\bf 5. Evaluation of the symplectic current.}
\medskip

In order to evaluate the symplectic current a little more work
is required. To start with (as in the preceeding work~\cite{Ca03})
it is convenient to go over from parallel variations to the corresponding
Lagrangian variations using the relations
\be
\dP x^\mu_{, \ii}=x^\nu_{,\ii}\nabla_{\!\nu}\xe^\mu \,
,\label{56-}
\ee
and
\be
\dP{\Tred}_\mu{^{\nu}}=\dL{\Tred}_\mu{^{\nu}}+
{\Tred}_\rho{^{\nu}}\nabla_{\!\mu}\xe^\rho-{\Tred}_\mu{^{\rho}}
\nabla_{\!\rho}\xe^\nu\, .\label{57-}
\ee
The vectorial version (\ref{42-}) of the symplectic current is
thereby obtained in the form
\be
\omegar^\nu=\Vert\gamme\Vert^{-1/2} x^\nu_{\, ,\ii}\,
\dP\pred_\mu^{\ \ii}\wedge\xe^\mu\, ,\label{58-}
\ee
with
\be \Vert\gamme\Vert^{-1/2} x^\nu_{\, ,\ii}\,\dP
\pred_\mu^{\ \ii }=\dL\Tred_{\!\mu}^{\ \nu}-\Tred_{\!\rho}^{\ \nu}
\nabla_{\!\mu}\xe^\rho+\Tred_{\!\mu}^{\
\nu}\overline\nable_{\!\rho} \xe^\rho\, .\label{59-}
\ee

The advantage of Lagrangian variations is their convenience for
relating the relevant intrinsic physical quantities via the
appropriate equations of state. Following the example of  Friedman
and Schutz~\cite{FrSc75}  in the context of ordinary relativistic
fluid dynamics, we use the second order derivative of the action
with respect to the background metric to obtain the hyper Cauchy
tensor (generalised elasticity tensor) according to the
prescription
\be
{\Cred}{^{\mu\nu\rho\sigma}}=\Vert\gamme\Vert^{-1/2}{\partial\big(
\Tred^{\mu\nu}\Vert\gamme\Vert^{1/2} \big)\over \partial
g_{\rho\sigma}} ={\Cred}{^{\rho\sigma\mu\nu}}\, .
\label{60-}
\ee
The Lagrangian
variation of the surface stress energy tensor will thus be
obtained in the form
\be \dL\Tred^{\mu\nu}= \big({\Cred}{^{\mu\nu\rho\sigma}}-{_1\over^2}
\Tred^{\mu\nu}\ete^{\rho\sigma}\big)\,\dL g_{\rho\sigma}\,
.\label{61-}\fe The symplectic current can thereby be expressed in
purely tensorial form as
\be \omegar^\nu=\big(2{\Cred}{_{\mu\ \rho}^{\ \, \nu\ \sigma}}
\overline\nable_{\!\sigma}\xe^\rho+{\Tred}{^{\nu\rho}}
\overline\nable_{\!\rho}\xe_\mu\Big)\wedge \xe^\mu\, . \label{62-}
\fe Using the explicit expression (\ref{12-}) for the surface
stress energy tensor in a generic elastic string model, the
corresponding expression for the required hyper Cauchy tensor is
obtainable from the definition (\ref{60-}) in the form
%
$$
{\Cred}{^{\mu\nu\rho\sigma}}=\Lred\big({_1\over^2}\ete^{\mu\nu}
\ete^{\rho\sigma}-\ete^{\mu(\rho}\ete^{\sigma)\nu}\big)
+{\calK\over 2}\big(\ete^{\mu\nu}\cur^\rho \cur^\sigma\!
+\!\ete^{\rho\sigma}\cur^\mu\cur^\sigma-4\cur^{(\mu}\ete^{\nu)(\rho}
\cur^{\sigma)}\big)$$
\be + \; \; \calK^2{{\rm d}\calK\over {\rm
d}\vw}\cur^\mu\cur^\nu \cur^\rho\cur^\sigma\, .\hskip 5 cm
\label{63-}
\ee

\bigskip\noindent
{\bf 6. The chiral case}
\medskip

As remarked above, the only way in which the off shell action for
the chiral model differs from that of the generic elastic case is
that instead of being held constant the auxiliary field \bb
\vpsi\fb is treated as a free variable. However, as its gradient
is not involved in the action, this extra variable will not give
rise to any corresponding momentum contribution, so the formulae
of the two preceding sections will remain valid for the chiral
model as characterised on shell by the current nullity condition
(\ref{17-}) and the coresponding restrictions (\ref{15-}).

It was remarked that for a given value of the overall
normalisation as fixed by the mass scale $\mred$, the same unique
chiral model with the same on shell stress energy tensor
(\ref{16-}) is obtained independently of the equation of state.
Notice, however, that due to the presence of the final term
proportional to ${\rm d}\calK/{\rm d}\vw$, the formula (\ref{63-})
for the hyper Cauchy tensor gives a result that {\it does} depend
on the choice of equation of state even in the chiral limit for
which \bb\vw=0\fb and $\calK=1$. The simplest possibility is
provided by the choice~\cite{CaPe99} of Witten's simple linear
equation of state  (\ref{22-}), which gives ${\rm d}\calK/{\rm
d}\vw=0$, so that the final term in (\ref{63-}) will drop out
altogether.  However, for other choices such as the more useful
one (\ref{23-}) discussed in section 2, there will be an extra
term proportional to $\cur^\mu\cur^\nu\cur^\rho\cur^\sigma$ with
an arbitrary proportionality constant. Although it is uniquely
defined on shell, the reason why the chiral model does not provide
a unique specification of the on shell hyper Cauchy tensor is that
the range of variation in the relevant partial derivative formula
(\ref{61-}) is restricted by the requirement that, in order to
preserve the nullity condition (\ref{17-}), the allowable
displacements must be such as to ensure that the Lagrangian metric
variation satisfies the on shell chiral variation condition \be
\cur^\rho\cur^\sigme\,\dL g_{\rho\sigma}=0\, . \label{65-} \ee
However, for this same reason, the final term in (\ref{63-}) will
provide no contribution to the on shell value of the corresponding
symplectic tensor, whose value for the chiral model will thus be
given unambiguously by substitution in (\ref{62-}) of the
expression obtained from the linear action (\ref{22-}), namely \be
{\Cred}{^{\mu\nu\rho\sigma}}=\mred^2\big(\ete^{\mu(\rho}
\ete^{\sigma)\nu}-{_1\over^2}\ete^{\mu\nu}\ete^{\rho\sigma}\big)
+{_1\over ^2}\big(\ete^{\mu\nu}\cur^\rho \cur^\sigma\!
+\!\ete^{\rho\sigma}\cur^\mu\cur^\sigma-4\cur^{(\mu}\ete^{\nu)(\rho}
\cur^{\sigma)}\big)\, . \label{66} \ee In terms of the orthogonal
projector  \bb\perpe^{\!\mu}_{\,\nu}= g^\mu_{\ \nu}-\ete^\mu_{\
\nu}\fb, it can be seen that this leads to an expression giving
the conserved symplectic current for the chiral string model in
the form \be \omegar^\nu=\big(\cur_\mu\cur_\rho\ete^{\nu\sigma} +
2\Tred_\mu^{\, [\sigma}\ete_\rho^{\,\nu]}
-\Tred^{\nu\sigma}\perpe_{\mu\rho}\big)
\xe^\mu\wedge\overline\nable_{\!\sigma}\xe^\rho\, , \ee in which
the chiral stress energy tensor is given by (\ref{16-}).

\bigskip
\noindent {\bf {Conclusion}}
\medskip

Our aim in this paper was first and foremost to calculate the
symplectic current central to the covariant canonical analysis of
Witten and others~\cite{Wi86,CrWi87,Zu87,So94,CF98,Nu00,Ro02} for
generic elastic and chiral string models.  The purpose of this
analysis was as a precursor to a covariant quantization.

Our work is based on the concise and efficient covariant analysis
developed in~\cite{Ca93,BaCa95,BaCa00}, rather than the more cumbersome frame
dependent methods which have been used by others in the case of
the simpler NG string. Our results are presented in sections 5 and
6 for the elastic and chiral strings respectively. In the process,
in section 2a and the appendix, we also studied with care the
different formulations of the chiral string model.  We showed the
equivalence of the Witten action (used in most studies to date),
with the square root action (equation (\ref{23-})) and in this
way we were able to understand in a different way the integrability
properties of the chiral string.

\newpage

\bigskip
\noindent
{\bf {Acknowledgements}}
\medskip

D.Steer would like to thank M.~Ruiz-Altaba for many discussions on
the quantisation of chiral strings and the appendix of this
article.

\bigskip
\noindent
{\bf {A. Appendix}}
\medskip

In this appendix we demonstrate how the Polyakov-action
(\ref{Lpolyfinalt}) can be obtained from the original square root
action (\ref{23-}) using Dirac's Hamiltonian formalism for
contrained systems \cite{Dirac,Hansen}.
The bulk of this appendix is
valid for arbitrary $\vw$ and hence for general elastic string
models: the special chiral limit $\vw=0$ will be discussed after
equation (\ref{eq:chirrr}) where the Poisson algebra of
constraints --- including the chiral one --- is studied. (For a
different work on constrained superconducting membranes, see
\cite{CR2}.)

Our starting point is the action (\ref{23-})
\be
{\calI} = \int {\rm d}^2\sigme {\calL}
 = -\mred^2 \int {\rm d}^2\sigme  \Vert\gamme\Vert^{1/2}
\big(1+\vw/\mred^2\big)^{1/2}
\ee
in an arbitrary background metric $g_{\mu \nu}$ and brane embedding $x^{\mu}$.
Let the two world-sheet coordinates be denoted by
$\sigme^{(0,1)}\equiv(\tau,\sigme)$ with $\tau$ time-like,
and ${\,}'=\partial_1$ and $\dot{\,} =
\partial_0$.  The canonical momentum
densities $\Pred_{\mu} = \delta {\calL}  /\delta \dot{x}^{\mu}$,
$\Pired = \delta {\calL}  /\delta \dot{\vphi}$ and $\Pred_{\vpsi}
= \delta {\calL} / \delta \dot{\vpsi} = 0$ are given by
\ba
\Pred_{\mu} &=& \frac{\mred^4}{{\calL}} \left[ (\dot x \cdot x')
x^{\prime}_{\mu} -x^{\prime2} \dot{x}_\mu +
\frac{\vpsi^2}{\mred^2} \left( x_{\prime\mu} \dot{\vphi} \vphi' -
\dot{x}_{\mu} \vphi^{\prime2} \right) \right],
\\
\Pired &=& - \vpsi^2 \frac{\mred^2}{{\calL}} \left[ x^{\prime2}
\dot{\vphi} - (\dot{x} \cdot x') \vphi' \right]
\ea
and the phase space of the system is
$\{x^{\mu},\vphi,\Pred_{\mu},\Pired\}$.  It follows directly from
these definitions of $\Pred_{\mu}$ and $\Pired$ that there two are
primary constraints independent of $\dot{x}$ and $\dot{\vphi}$.
One (see equation (\ref{c2}) below) is straightforward to deduce.
In order to find the other, it is easiest express both $\Pred^2$
and $\Pired^2$ in terms of $\vw
=\vpsi^2\gamme^{\ii\ji}\vphi_{,\ii}\vphi_{,\ji} $ defined in
(\ref{3-}), leading to
\ba
-\frac{\vw}{\mred^6} \left[\Pred^2 + \vpsi^2 \vphi'^2 \mred^2
\right] &=& x'^2 + \frac{1}{\mred^4} \left[  \Pred^2 + 2 \vpsi^2
\vphi'^2 \mred^2 \right], \label{cc1}
\\
\vw \left[ x'^2 \vpsi^2 + \frac{\Pired^2}{m^2} \right] &=&
\vpsi^4 \vphi'^2 - \Pired^2. \label{cc2}
\ea
Eliminating $\vw$ between these equations gives the other primary
constraint, so that the two constraints are
\ba
\calH_1 &\equiv& \Pred^2 + \frac{\mred^2 \Pired^2}{\psi^2} + \mred^4
x'^2 + \mred^2 \vpsi^2 \vphi'^2 = 0, \label{c1}
\\
\calH_2 & \equiv & \Pred_{\mu} x'^{\mu} + \Pired \vphi' = 0.
\label{c2}
\ea
The Nambu-Goto limit is obtained when $\vpsi\to0$ (note that
$\Pired^2 \sim \vpsi^4$) in which case these expressions reduce to
the standard constraints \cite{Hansen} for the bosonic string.

Had we started instead from the linear Witten action (\ref{22-})
and calculated the new momenta, it would have followed that the
second constraint (\ref{c2}) still holds.  On the other hand, a
constraint of the first form linking $\Pred^2$ and $\Pired^2$ can
no longer be obtained: though one can still express $\Pred^2$ as a
function of $\vw$ in a manner analogous to (\ref{cc1}), there is
no analogue of (\ref{cc2}) for the linear Witten action.  Indeed
$\Pired^2$ is now depends on $\dot{\vphi}$ and $\dot{x}^{\mu}$ in
a combination that can no longer be expressed solely in terms of
$\vw$.

Going back to the square root action, it follows from the above expressions
that the canonical
Hamiltonian density vanishes
\be
{\calH}_c \equiv \Pired \dot{\vphi} + \dot{x}_\mu \Pred^{\mu} -
{\calL} = 0. \label{canonH}
\ee
As explained by Dirac \cite{Dirac,Hansen}, this Hamiltonian is
ambiguous since one is free to add arbitrary multiples of the
vanishing primary constraints and secondary constraints. These
secondary constraints arise from imposing that the primary
constraints should be preserved under time evolution \cite{Dirac}.
Here, given that ${\calH}_c=0$ and that the Poisson brackets of
the ${\calH}_{1,2}$ are closed (see equations
(\ref{com1})-(\ref{com3}) below), it follows that there are no
secondary constraints.  Hence the `total' Hamiltonian density
\cite{Dirac,Hansen} which determines the dynamics of the system is
simply a linear combination of the primary constraints,
\be
{\calH} \equiv {\calH}_c +\frac{\glambda}{2\mred^2} \calH_1+
\gmu\; \calH_2 = \frac{\glambda}{2\mred^2} \calH_1+ \gmu\;
\calH_2, \label{Htot}
\ee
where $\gmu$ and $\glambda$ are dimensionless Lagrange
multipliers.
We now briefly turn away from the phase space approach to
calculate the (Polyakov) Lagrangian
\be
\calL_P \equiv \dot{x}_{\mu} \Pred^{\mu} + \dot{\varphi} \Pired -
{\calH}
\label{Lpoly}
\ee
corresponding to (\ref{Htot}).
From the equation of motion $\dot{x}_\mu(\tau,\sigme') = \int d
\sigme \{ {x}_\mu(\tau,\sigme'),{\cal H}(\tau,\sigme) \}$ (with a
similar one for $\vphi$)
it follows that $\Pred_{\mu} =
(\mred^2/{\glambda })(\dot{x}_{\mu} - \gmu x'_{\mu})$, and
similarly $\Pired = - (\vpsi^2 / \glambda) (\gmu \varphi' - \dot{\varphi})$.
These enable all the momenta to be eliminated from $\calL_P$ of
equation (\ref{Lpoly}), and one finds in the particular case of the square
root action
\be
{\calI} = \int {\rm d}^2\sigme
{\calL}_{P} =
-\frac{\mred^2}{2} \int {\rm d}^2\sigme
\Vert\he\Vert^{1/2} \he^{\ii \ji} \left(g_{\mu \nu} \partial_{\ii}
x^{\mu} \partial_{\ji}x^{\nu} + \frac{\vpsi^2}{\mred^2}
\partial_{\ii}\vphi \partial_{\ji}\vphi \right)
\label{Lpolyfinal} \ee where the components of $\he_{ij}$ are
simply expressible in terms of $\glambda$ and $\gmu$: \be
\sqrt{-\he}\he^{00} = -\frac{1}{\glambda},  \qquad
\sqrt{-\he}\he^{01} = \frac{\gmu}{\glambda}, \qquad
\sqrt{-\he}\he^{11} = \glambda - \frac{\gmu^2}{\glambda}.
\label{hrelns} \ee This is the result given in
(\ref{Lpolyfinalt}).  Note that action (\ref{Lpolyfinal}) has
reparametrization and Weyl scale invariance generated by the two
primary constraints $\calH_{1,2}$, and reflected in the two free
parameters $\glambda$ and $\gmu$.  For instance, an appropriate
choice of coordinates can set $\he_{\ii \ji} = \eta_{\ii \ji}$
where $\eta_{\ii \ji}$ is the Minkowski metric (or more simply one
can choose $\glambda=1$ and $\gmu=0$ in (\ref{hrelns})). As for
the Nambu string \cite{Green}, this is one appropriate starting
point with which to quantize the chiral string \cite{Marti}.

As noted above, this result is a special property of the square
root action. For example, in the case of the linear (Witten)
action, equation (\ref{22-}), only the primary constraint
$\calH_2$ exists so that
the `full' Hamiltonian would now contain a single term,
$\calH=\gmu  \calH_2$.  Following a similar procedure to that
above shows that one would no longer be able to eliminate the
momenta in order to calculate the corresponding Lagrangian. (The
equations of motion do not give equations for $\Pred_{\mu}$ and
$\Pired$ but rather the equalities $\dot{x}_{\nu} = \gmu x'_{\nu}$
and $\dot{\vphi} = \gmu \vphi'$.)

We now make a comment about the algebra of constraints for the
square root action. As for
the Nambu string \cite{Hansen} (and also as previously noted), it
follows from the equations of motion that generators $\calH_1$ and
$\calH_2$ form a closed algebra (i.e.\ they are first class
constraints):
\ba
\{ \calH_1(\sigme), \calH_1 (\sigme')\} & \propto & \calH_2
\frac{\partial }{\partial \sigme} \delta(\sigme - \sigme'),
\label{com1}
\\
\{ \calH_1(\sigme), \calH_2 (\sigme')\} & \propto & \calH_1
\frac{\partial }{\partial \sigme} \delta(\sigme - \sigme'),
\\
\{ \calH_2(\sigme), \calH_2 (\sigme')\} & \propto & \calH_2
\frac{\partial }{\partial \sigme} \delta(\sigme - \sigme').
\label{com3}
\ea
Is it also consistent to impose the chiral constraint at the
algebraic level?  As noted in section 2, the chiral constraint
$\partial {\calL} /
\partial \psi=0$ imposes the nullity condition
\be
\vw =  0, \label{eq:chirrr}
\ee
which we would like to express on the same footing as the primary
constraints (\ref{c1}) and (\ref{c2}) --- that is, in terms of
momenta only. Let this constraint be denoted by $\calH_3$. From
equations (\ref{cc1}) and (\ref{cc2}), if therefore follows that
\be
\calH_3 \equiv \alpha \left[\mred^4 x'^2 + \Pred^2 + 2 \vpsi^2
\vphi'^2 \mred^2 \right] + \vpsi^2 m^2 \beta \left[ \vphi'^2 -
\frac{\Pired^2}{\vpsi^4} \right] = 0,
\ee
for initially arbitrary $\alpha$ and $\beta$.  Recall, however
(section 2), that the null constraint $\vw=0$ is consistent with
the dynamical equations, being conserved under time evolution. At
the level of constraints, we demand that the chiral constraint
$\calH_3$ should also be conserved under time evolution
--- in other words, that it does not lead to a secondary
constraint.  Equivalently this imposes that the Poisson brackets
of $\calH_3$ with the other two constraints $\calH_1, \calH_2$
should be closed.  In turn one can show that this is true only
for $\alpha=1/2$ and $\beta=-\alpha(1 \pm \sqrt{2})$ in which case
\ba
\{ \calH_3(\sigme), \calH_1 (\sigme')\} & \propto & \calH_2 \frac{\partial }{\partial \sigme} \delta(\sigme - \sigme'),
\\
\{ \calH_2(\sigme), \calH_3 (\sigme')\} & \propto & \calH_3 \frac{\partial }{\partial \sigme} \delta(\sigme - \sigme'),
\\
\{ \calH_3(\sigme), \calH_3 (\sigme')\} & \propto & {\calH_2} \frac{\partial }{\partial \sigme} \delta(\sigme - \sigme').
\ea

\end{document}